\documentclass[useAMS,usenatbib]{mn2e}

\usepackage{graphics}
\usepackage{graphicx}

\title[Constraint on a variation of $\mu$ from H$_2$
absorption]{Constraint on a variation of the proton-to-electron mass
ratio from H$_2$ absorption towards quasar
Q2348$-$011}
\author[J. Bagdonaite et al.]{Julija Bagdonaite$^{1}$\thanks{E-mail:
j.bagdonaite@vu.nl; mmurphy@swin.edu.au; L.Kaper@uva.nl; w.m.g.ubachs@vu.nl},
Michael T. Murphy$^{2}$\footnotemark[1], Lex
Kaper$^{1,3}$\footnotemark[1] and Wim Ubachs$^{1}$\footnotemark[1]
\\ 
$^{1}$Department of Physics and Astronomy, LaserLaB, VU University, De Boelelaan
1081, 1081 HV Amsterdam, The Netherlands
\\
$^{2}$Centre for Astrophysics and Supercomputing, Swinburne University of
Technology, Victoria, 3122, Australia
\\
$^{3}$Astronomical Institute Anton Pannekoek, Universiteit van Amsterdam,
1098 SJ Amsterdam, The Netherlands}
\begin{document}

\date{Accepted 2011 December 1.  Received 2011 December 1; in original form 2011 November 18}

\pagerange{\pageref{firstpage}--\pageref{lastpage}} \pubyear{2011}

\maketitle

\label{firstpage}

\begin{abstract}
Molecular hydrogen (H$_2$) absorption features observed in the line-of-sight to
Q2348$-$011 at redshift z$_{\rm abs} \simeq $2.426 are analysed for the purpose
of detecting a possible variation of the proton-to-electron mass ratio $\mu\equiv
m_p/m_e$.  By its structure Q2348$-$011 is the most complex analysed H$_2$
absorption system at high redshift so far, featuring at least seven distinctly
visible molecular velocity components. The multiple velocity components associated
with each transition of H$_2$ were modeled simultaneously by means of a
comprehensive fitting method. The fiducial model resulted in  $\Delta\mu/\mu
=(-0.68\pm2.78) \times 10^{-5}$, showing no sign that $\mu$ in this particular
absorber is different from its current laboratory value. Although not as tight a
constraint as other absorbers have recently provided, this result is consistent
with the results from all previously
analysed H$_2$-bearing sight-lines. Combining all such measurements yields a
constraint of $|\Delta\mu/\mu|$ $\le 10^{-5}$ for the redshift
range $z =$ (2--3).

\end{abstract}

\begin{keywords}
methods: data analysis, quasars: absorption lines, cosmology: observations
\end{keywords}

\section{Introduction}

The spectrum of molecular hydrogen is known to be a testing ground to search for
temporal variation of the proton-electron mass ratio, $\mu\equiv
m_p/m_e$, on cosmological time-scales \citep{thompson}. The spectral lines of the Lyman and
Werner absorption bands of H$_2$ shift into an atmospheric transmission window
($\lambda >3050$ $\textrm\AA$) for absorption systems at redshift $z \ga 2$, and
thus become observable with ground-based optical telescopes. Wavelengths of H$_2$
spectral lines as observed in high redshift absorption systems are compared to
the wavelengths measured in the laboratory for which now very accurate
calibrations exist (\citealt{philip2004}; \citealt{salumbides};
\citealt{bailly}) as well as for the HD isotopologue
(\citealt{ivanov2008}; \citealt{ivanov2010}), which is currently also detected
at high redshift. Values for the sensitivity coefficients, $K_i$, expressing the shift of each spectral
line as a result of a drifting $\mu$, have been established to sufficient accuracy for
H$_2$ (\citealt{meshkov2006}; \citealt{ubachs2007}). Hence a condition is
accomplished that searches for cosmological $\mu$ variation via H$_2$ spectra
solely depend on the accuracy of the astrophysical data. 

From the large number of damped Lyman-$\alpha$ systems (DLAs) identified only
some 20 are known to harbor detectable H$_2$ spectral features
(\citealt{ledoux2003}; \citealt{srianand2005}; \citealt{ubachs2007}). As was
discussed recently \citep{ubachs2011} only a handful of those high redshift
absorption systems can deliver an H$_2$ spectrum of sufficient quality, \emph{i.e.} a
reasonable signal-to-noise (S/N) ratio on the continuum level to be obtained in
reasonable data collection times at large-dish telescopes, and a large number of
H$_2$ transitions ($\ga$ 50). The high quality spectra of the Q0347$-$383 and
Q0405$-$443 systems obtained with the ESO Very Large Telescope (VLT), equipped
with the high resolution Ultraviolet and Visible Echelle Spectrograph (UVES)
initially yielded an indication for a possible drift in $\mu$, which is
expressed via $\Delta\mu/\mu \equiv (\mu_{z} - \mu_{\rm
lab})/ \mu_{\rm lab}$ \citep{reinhold2006}. A more
sophisticated re-analysis of the same spectra by the so-called comprehensive
fitting method, also invoking an improved thorium-argon (Th-Ar) calibration,
reduced the initial finding of a 4-$\sigma$ effect of a positive $\Delta\mu/\mu$ to a
1.5$-$2.0-$\sigma$ effect \citep{king2008}.
The same spectra were also re-analysed by \citet{wendt2008} and by
\citet{thompson2009}, while \citet{wendt2010} re-observed Q0347$-$383 with
VLT-UVES to find a 1.5-$\sigma$ effect. The highest quality system observed so
far is J2123$-$0050. Observations with both the Keck Telescope \citep{malec2010} and
VLT \citep{weerdenburg2011} were incorporated in a $\mu$-variation analysis. The two
studies yielded constraints on $\Delta\mu/\mu$ which are in very good agreement
with each other. The averaged constraint for J2123$-$0050 is $\Delta\mu/\mu =
(7.6\pm3.5)\times10^{-6}$. The system Q0528$-$250 has been at the heart of
$\mu$-variation analyses since the first attempts by \citet{varshlovich1993} to
derive a constraint on $\Delta\mu/\mu$ from a low-resolution spectrum. \citet{king2008}
derived a tight constraint from a comprehensive fitting analysis of a
VLT-spectrum of Q0528$-$250; a recent re-observation of this high-quality
system, again at VLT, confirmed the previous conclusion of a tight constraint on
$\Delta\mu/\mu$ \citep{king2011}. Based on the combined results derived
from high redshift H$_2$ absorbers it can be concluded that $|\Delta\mu/\mu| < 1 \times 10^{-5}$ for
redshifts $z=2-3$.

The present study presents a detailed analysis of the search for $\mu$-variation
of an absorber system toward the quasar Q2348$-$011. The physical conditions of
this system had been investigated before based on data from the VLT (\citealt{petitjean2006};
\citealt{ledoux2006}; \citealt{noterdaeme2007}). Q2348$-$011 was shown to be an
exceptional sight-line, with a number of DLAs and sub-DLAs
present, and a very complex velocity structure in the major H$_2$ absorbing
system at $z \simeq 2.426$. At least seven distinctly visible
absorption features are associated with each H$_2$ transition in this system,
while in other quasar sight-lines H$_2$ absorption (if detected) exhibits much
simpler profiles consisting typically of one or two features. For the present study
dedicated observations were performed at VLT focusing on
improvement of S/N and on wavelength calibration of the spectrum. A preliminary
$\mu$-variation analysis of this absorber was presented in Ubachs et al. (2011).
The current work improves upon the analysis but the constraint on $\Delta/\mu$
is of similar precision as presented previously. It
is demonstrated that a complex velocity structure consisting of at least seven
distinct H$_2$ velocity features can be disentangled by the method of
comprehensive fitting (\citealt{king2008}; \citealt{malec2010};
\citealt{king2011}). 

\section[]{Data}

The new spectrum of the quasar Q2348$-$011 was obtained with the UVES
spectrograph on the ESO VLT on four consecutive nights (2008
August 18--21). Q2348$-$011 is not a
particularly bright object (R = 18.31) so a long exposure time was required.
Fifteen exposures were taken, making up a combined total of 19.25 hours of
observation. After each science exposure, without any intervening grating resets, a Th-Ar exposure was recorded
for calibration purposes. Seeing was in the range between 0$\farcs$53 and
2$\farcs$32. Besides the exposures noted above, the final spectrum incorporates
data obtained from the ESO data archive (observations from 2003, October
29--30). They contribute additional 4.50 hours but do
not have the individually taken Th-Ar calibration frames.

The raw 2D exposures were bias-corrected, flat-fielded, and the quasar flux
extracted using the Common Pipeline Language version of the UVES
pipeline\footnote{
http://www.eso.org/observing/dfo/quality/UVES/pipeline/pipe $\_$reduc.html.}.
The wavelength calibration was established by extracting each Th-Ar frame using
the object profile weights from its corresponding quasar exposure. Further
details of the wavelength calibration procedure are described in
\citet{murphy2007}. The root-mean-square wavelength calibration residuals were
typically $\approx$70~m~s$^{-1}$ in the Lyman-$\alpha$ forest portion of the spectrum,
where the H$_2$ lines fall. The wavelength scale of each quasar exposure was
corrected from air to vacuum, and to the heliocentric reference frame, and the
flux was re-dispersed on to a common log-linear scale before being co-added,
using {\sevensize UVES$\_$POPLER}\footnote{Developed and maintained by M.~Murphy;
see http://astronomy.swin.edu.au/$\sim$mmurphy/UVES$\_$popler.}, a code
specifically written to re-disperse and combine reduced exposures from UVES.

The spectrum of Q2348$-$011 covers wavelengths 3572--9467 $\textrm\AA$, with
gaps at 4520--4621 and 7505--7665 $\textrm\AA$. The final spectrum has a
resolving power of  R $\sim$ 57,000 and 63,000, and the log-linear dispersion is set to
2.5~km~s$^{-1}$ and 1.5~km~s$^{-1}$ in the blue ($<4000~{\textrm \AA}$) and in
the red ($>4000~{\textrm \AA}$) parts,
respectively. CCD pixels were binned by a factor of 2 in both spatial and
spectral directions for all exposures in the blue, and no binning was applied
for the red exposures. In the blue part where H$_2$ features are detected the
average S/N ratio is about 25.

\section{The comprehensive fitting method}
\label{section:method}
The main aim of the present analysis is to determine the spectral positions of
the H$_2$ absorption features present in a quasar spectrum as accurately as possible
since their overall pattern defines the value of \(\Delta\mu / \mu\): each
absorption line of the H$_2$ Lyman and Werner bands shows a unique shift for a
given change of \(\mu\) that depends on the vibrational and rotational quantum numbers of the
upper and lower energy states. The shifted wavelength
\(\lambda_i\) is related to the rest wavelength \(\lambda_0\) by
\begin{equation} 
\lambda_i = \lambda_0(1 + z)(1+ K_i
\frac{\Delta\mu}{\mu}) 
\label{eq1}
\end{equation}
where \(z\) is the redshift of the absorber, and $K_{i}$ is the sensitivity
coefficient, different for each line. To achieve
this goal a so-called comprehensive fitting method can be applied. This method
requires fitting the absorption profiles of multiple H$_2$ transitions
simultaneously. It relies on a physical assumption that
all transitions arise from the same cloud (collection of clouds) of molecular
hydrogen and, therefore, they share the parameters describing the properties of the
clouds (see e.g., \citealt{malec2010}). 

Each H$_2$ transition can be detected in one or several absorption features
depending on how many molecular hydrogen clouds are penetrated by the light of the quasar. We
refer to each fitted component as a velocity component, as they are situated closely
in velocity space. It is assumed that a given velocity component has the same redshift, $z_{\rm abs}$, and Doppler
width, $b$, in all transitions independent of the rotational level, $J$.
Each $J$ level has a different population, represented by the column density
$N$, but for a given velocity component it is the
same for all transitions from the same $J$ level. To analyse the spectrum we use
the Voigt profile fitting program {\sevensize
VPFIT}\footnote{http://www.ast.cam.ac.uk/$\sim$rfc/vpfit.html.}, which permits
to tie the free parameters between various transitions. The optimal model is
found by fitting all suitable transitions at the same time.

The differential shifts of the molecular hydrogen lines are only known to be
possible through the \(\Delta\mu / \mu\) parameter (see Eq.~\ref{eq1}). It is
added as a free parameter after all other parameters have been optimized. The comprehensive
fitting approach allows minimization of the number of free parameters in the fit
and, thus, the reliability of the absorption model which eventually
translates to the robustness of \(\Delta\mu / \mu\) resulting from the model.

\section[]{Analysis}
\subsection{The absorption system}

\begin{figure}
  \centering
  \includegraphics[scale=0.85]{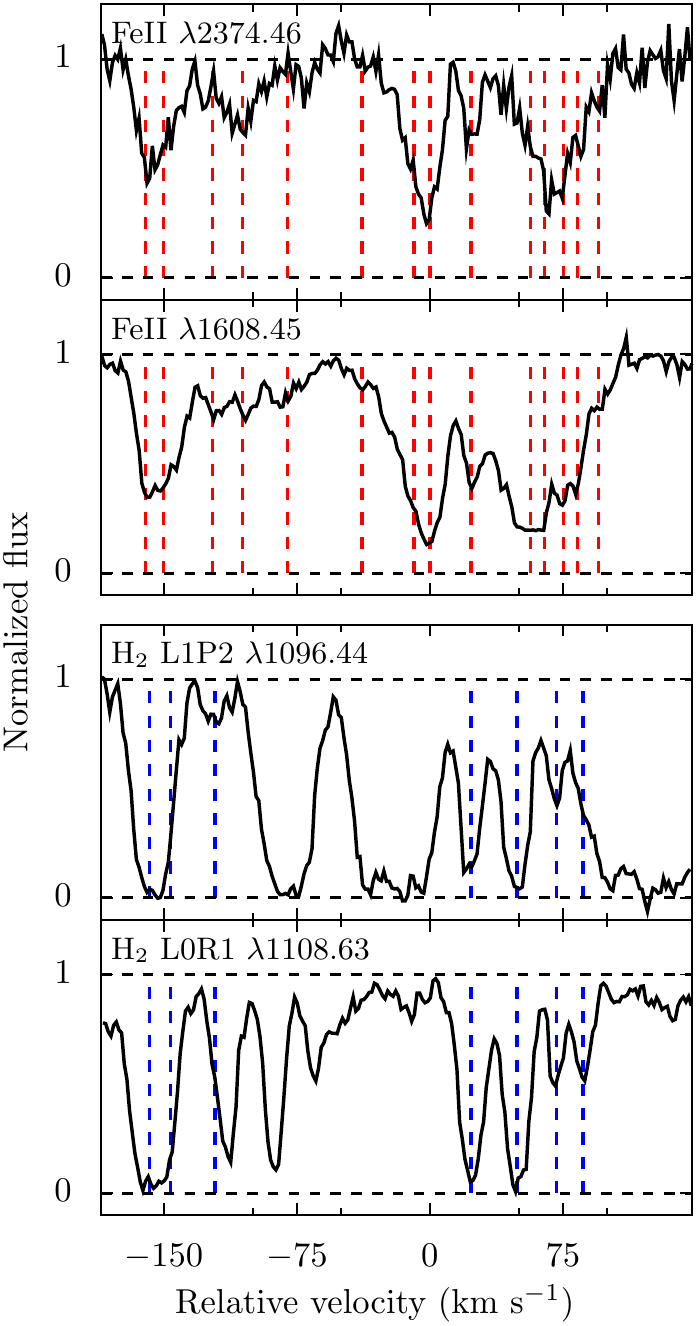}
 \caption{Two H$_2$ and two Fe{\sevensize~II} transitions are displayed on a
common velocity scale, centered at $z_0=2.4263230$. In the H$_2$ velocity
profile 7 distinct features are visible, while the Fe{\sevensize~II} profiles
consist of at least 14. Note that the third velocity component from the left in
the H$_2$ L0R1 profile, and the right-most component in the H$_2$ L1P2 profile
are blended with other intervening transitions in Ly-$\alpha$ forest. Due to
such blends, the signal of at least 7 absorption components is evident only when
multiple H$_2$ transitions are explored simultaneously.}
 \label{fig1}
\end{figure}

The wavelengths covered by the VLT-UVES spectrum provide 58 H$_2$ transitions
for rotational levels $J = 0-5$. All the observed lines arise from the Lyman
band. In Fig.~\ref{fig1} a couple of H$_2$
transitions are displayed. No lines of HD were detected in this system. In the
ideal case this spectrum would provide a sample of \(58\times7\) molecular hydrogen
absorption lines. However, not all of them are suitable
for the analysis. First of all, the broad Ly-\(\beta\) of an additional DLA at
$z \simeq 2.62$ falls in the wavelength range where H$_2$ lines are detected, namely near
3710 $\textrm\AA$. The break at $\le$3590 $\textrm\AA$ might be produced by
another absorber at $z \simeq 2.93$. These features make some of the H$_2$
transitions unavailable, and strongly damp some others. Also, the
weaker H{\sevensize~I} absorbers of the Ly-\(\alpha\) forest
contaminate many of the regions with H$_2$ lines.
Generally, the sight-line of the Q2348$-$011 is rich in absorbing systems. The
full list of detected metal absorbers is given in Table~\ref{table1}. Some of
the absorbers have their metal absorption lines overlapped with H$_2$ lines. The
metal absorption profiles are constrained by fitting their counterparts in the
red part, e.g., Fe{\sevensize~II} $\lambda$1096 of the $z\simeq 2.426$ absorber
can be constrained by tying it with Fe{\sevensize~II} $\lambda$1608. 

\begin{table}
    \caption{A list of all the confirmed detections of metal absorbers in the
spectrum of Q2348$-$011. Species whose transitions overlap some of the fitted
H$_2$ regions are in bold.}
    \centering
    \begin{tabular}{c c}
    \hline\hline
     Redshift & Species  \\ [0.5ex] \hline
     0.774 & Mg{\sevensize~II} \\
     0.863 & Mg{\sevensize~II}, Ca{\sevensize~II} \\
     {\bf 1.444} & {\bf C{\sevensize~IV}}, Mg{\sevensize~II} \\
     {\bf 2.426} & C{\sevensize~IV}, Si{\sevensize~IV}, N{\sevensize~V},
C{\sevensize~I}, C{\sevensize~I*}, P{\sevensize~II}, \\
	   & Zn{\sevensize~II}, Al{\sevensize~III}, {\bf Fe{\sevensize~II}},
S{\sevensize~II}, {\bf Si{\sevensize~II}} \\
     2.582 & C{\sevensize~IV}, Si{\sevensize~IV}, Si{\sevensize~II} \\
     {\bf 2.615} & C{\sevensize~IV}, Si{\sevensize~IV}, Fe{\sevensize~II},
Ni{\sevensize~ II}, Cr{\sevensize~II}, Al{\sevensize~II}, \\
           & {\bf Si{\sevensize~II}}, {\bf O{\sevensize~I}}, N{\sevensize~I} \\
     {\bf 2.739} & C{\sevensize~IV}, {\bf O{\sevensize~I}} \\
     {\bf 2.929} & C{\sevensize~IV}, {\bf Si{\sevensize~IV}} \\
     2.969 & C{\sevensize~IV}, N{\sevensize~V}\\

    \hline
    \end{tabular}
    \label{table1}
\end{table}  

\begin{figure}
  \includegraphics[scale=0.75]{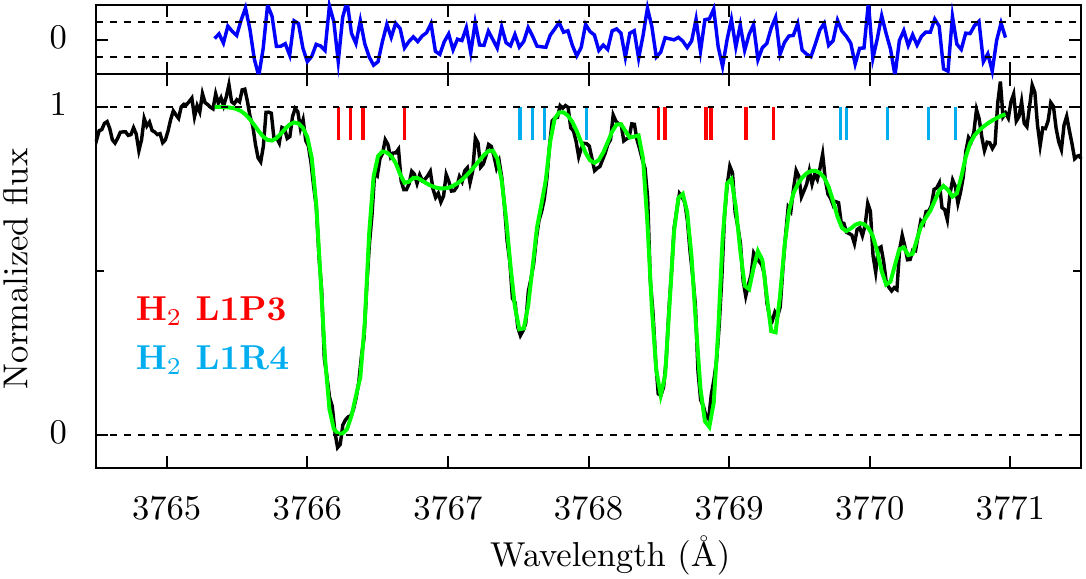}
 \caption{H$_2$ absorption profiles of two transitions: L1P3, and L1R4,
resulting from the fiducial model. Since the absorption arises in multiple
widely spread components (indicated by the red tick marks for L1P3, and by the blue
tick marks for L1R4), the two profiles overlap. The upper graph shows the 
normalized residuals of the fit; the dashed
lines mark $\pm1~\sigma$ limits. }
 \label{fig2}
\end{figure}

Since the velocity structure of H$_2$ consists of at least seven velocity
features spread over \(\sim\) 250~km~s$^{-1}$, nearby lying transitions exhibit
mutual overlap of velocity components. The self-blending of the H$_2$
transitions was often the decisive factor when setting the boundaries of usable
fitting regions (i.e.~regions over which $\chi^2$ was calculated): only rarely
did they include just one transition (see Fig.~\ref{fig2} as
an example).

\begin{figure*}
 \centering
  \includegraphics[scale=0.75]{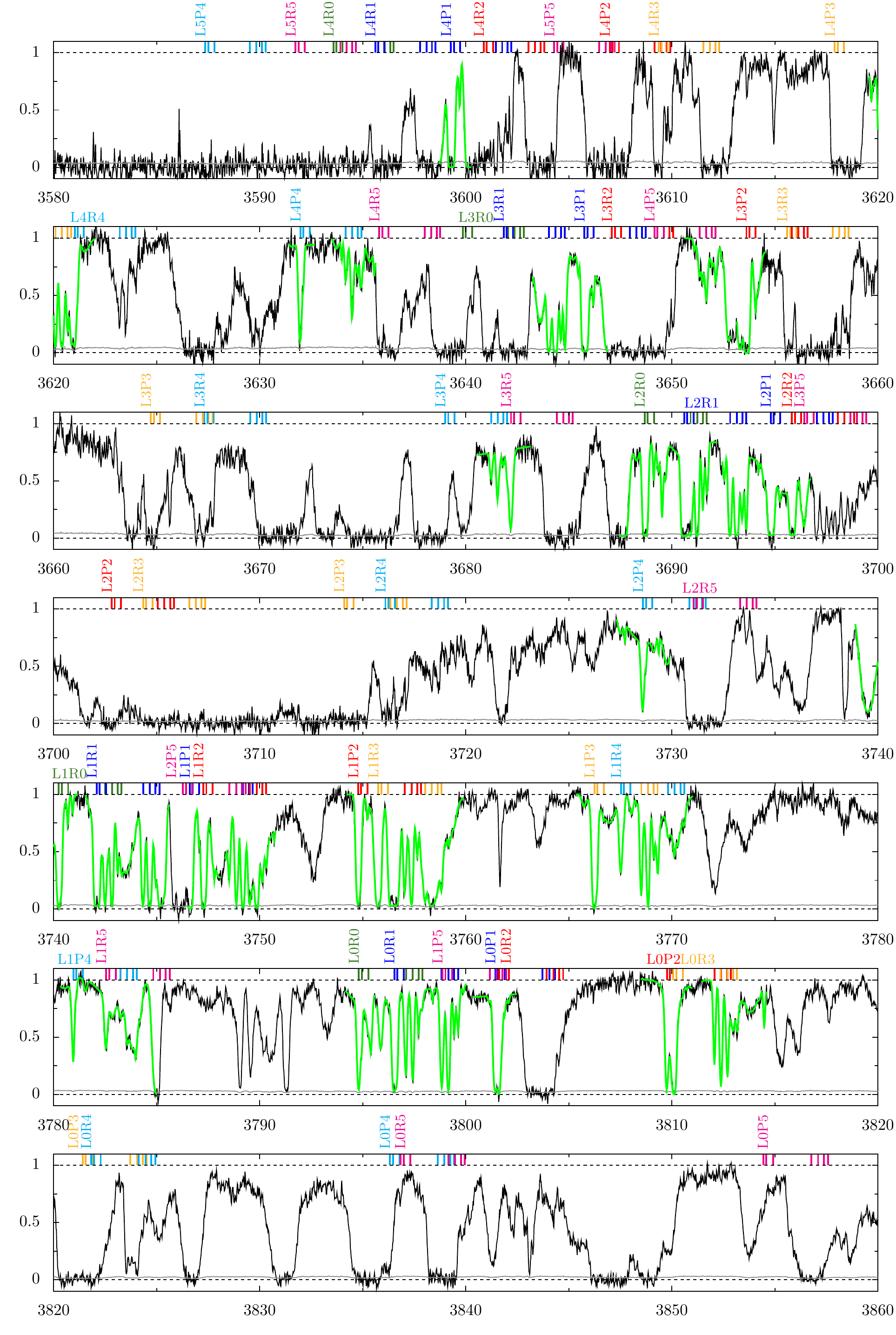}
 \caption{The part of the Q2348$-$011 spectrum containing the H$_2$ transitions.
Transitions from different $J$ levels are labeled in different colours. The
associated ticks show positions of the 7 (visually distinguishable) velocity
components. The broad absorption feature at 3710 $\textrm\AA$ is the
Lyman-$\beta$ line of the H{\sevensize~I} absorber at $z\simeq2.62$. The break
at the shortest wavelengths is due to another H{\sevensize~I} absorber at $z\simeq
2.93$. The green line shows the fitted regions. }
  \label{fig3}
\end{figure*}

Twenty-three regions with 32 different molecular hydrogen
transitions and eight regions with relevant metal absorption profiles were
selected for further analysis.
Although the total number of H$_2$ transitions available in this
spectrum is reduced due to blends with H{\sevensize~I} lines, the scarcity of
information is at least partially
compensated for by the presence of multiple velocity components associated
with each transition. Altogether, each transition contributes a higher
information content when it is imprinted on several velocity components.
A part of the spectrum with all available and all fitted H$_2$ transitions
indicated is provided in Fig.~\ref{fig3}.

\subsection{Exploring the velocity structure of H$_2$}
\label{section:exploring}

Seven absorption features can be distinguished by eye in the velocity structure
of molecular hydrogen in the Q2348$-$011 spectrum (see Fig.~\ref{fig1}, also
\citet{noterdaeme2007}). However, as it was shown by
\citet{murphy2008}, a single spectral feature might need to be
modeled by more than one velocity component so it is plausible that in
the case of Q2348$-$011 more than seven velocity components are actually
required to produce a model which is both physically realistic and statistically
acceptable. If a single Voigt profile is fitted to a feature which seems single
but another weaker blending line is actually present, then the centroid returned
from this fit will be shifted towards the blending line. The centroid wavelength
of the fitted feature might be expressed as the intensity-weighted mean of the
two blended lines. The positions of the lines play a crucial role in
$\mu$-variation analysis and this kind of underfitting may affect the results
and therefore should be avoided.

The approach to reaching the optimal model of H$_2$ absorption is to keep
adding velocity components until a statistically acceptable fit is achieved
(e.g., \citealt{malec2010}; \citealt{king2011}). 
The initial absorption model included seven velocity components of H$_2$. Voigt
profile parameters ($z$, $b$, $N$) of corresponding components are tied as
described in Section~\ref{section:method}. Each of the included regions
was revisited and refined many times until an adequate fit was reached. By
adequate we mean that the normalised residuals in each of the fitted regions are
not too large or too small (i.e. they are distributed around
[$-1$-$\sigma$:1-$\sigma$]), and no long
range correlations are seen in them. Finally, after the fit was optimized, a
value for $\Delta\mu/\mu$ was determined.

In a subsequent step an additional component was added to the left side of the
absorption profile (at
$\approx-150$ km~s$^{-1}$; see Fig.~\ref{fig1}). The reduced chi-square,
$\chi^{2}_{\nu}$, of this model is slightly better than that of the initial one with 7 velocity
components. The procedure of adding more components was continued further until they
started to be rejected from most $J$ levels ($>$12
component models). We note that several components get rejected from $J=$~4
and 5 transitions in all models. Generally, it is not unusual that weak
components are omitted from the lowly populated $J$ levels.
The higher $J$ levels are less populated than the lower $J$ levels, so weak
components of $J=$~4 and 5 transitions may not be detected. In addition, the
Q2348$-$011 spectrum provides only a few useful regions with $J=$~4 and 5
transitions, which is not enough to amplify the signal of the weak components.

It should be emphasized that $\Delta\mu/\mu$ is always introduced as a single
additional parameter after all the other free parameters of the given model have been optimized. This
ensures that $\Delta\mu/\mu$ is not biased away from zero in the process of
determining which velocity structure is statistically preferred. On the other
hand, it could mean that $\Delta\mu/\mu$ might be slightly biased \emph{towards}
zero because the preferred velocity structure might have `fitted away' some of
the relative shifts between transitions which a non-zero $\Delta\mu/\mu$ would
have caused. However, given that many H$_2$ transitions are fitted
simultaneously, the bias towards zero will be very weak.

\begin{figure}
 \centering
  \includegraphics[scale=0.8]{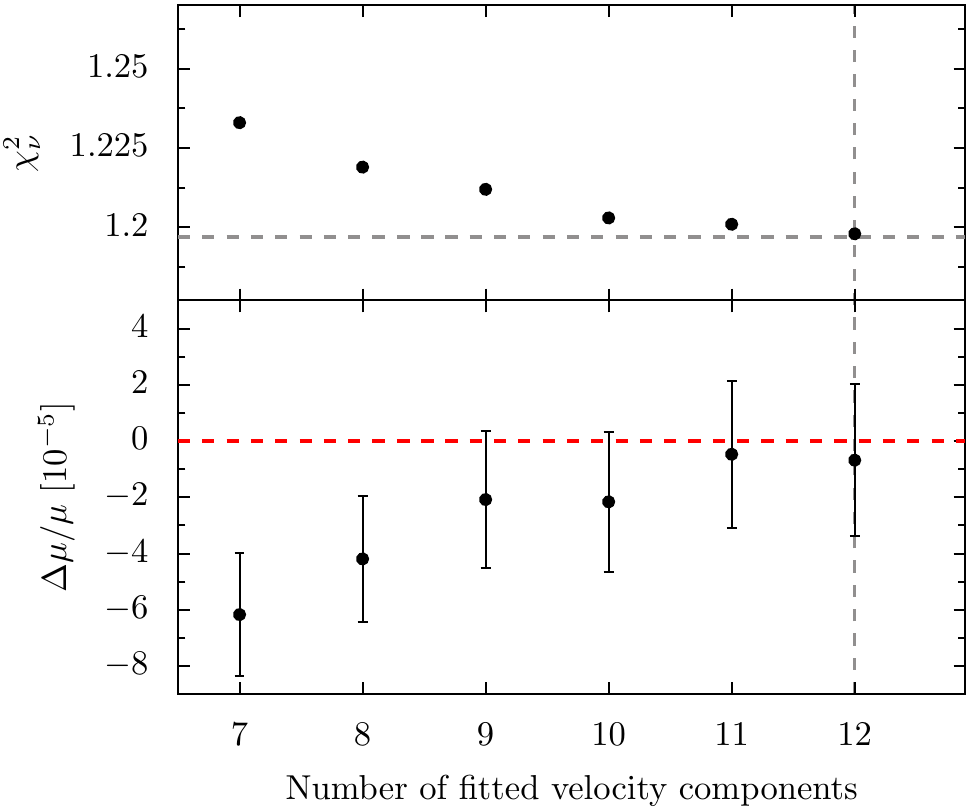}
 \caption{Lower graph: $\Delta\mu/\mu$ values as determined from various models.
Error bars indicate estimated 1-$\sigma$ statistical uncertainties. Upper graph:
$\chi^{2}_{\nu}$ of the fit decreases as more components are added. A
model with 12 components has the lowest $\chi^{2}_{\nu}$. Adding more components
caused inconsistencies in the fit.}
 \label{fig4}
\end{figure}

Results of the various fitting runs are displayed in Fig.~\ref{fig4},
together with the goodness-of-fit measure, $\chi^{2}_{\nu}$.
As it can be seen in Fig.~\ref{fig4}, the $\chi^{2}_{\nu}$ monotonically
decreases when more components are added. For models with 10, 11 and 12
components the difference in $\chi^{2}_{\nu}$ is very small though. This is also
true for the derived $\Delta\mu/\mu$ values -- they match well within
1-$\sigma$. For further analysis the model with 12 velocity components has been adopted as fiducial.

\subsection{Robustness of the model}
\label{tests}

When all available transitions are included in the 12 component model, the fit
delivers $\Delta\mu/\mu = (-0.68\pm2.70_{\rm stat}) \times 10^{-5}$. Next the
fiducial model is tested for its robustness. In the
process of creating the model, many assumptions and decisions have been made,
e.g., how parameters are tied between transitions and which regions are fitted.
It can be tested how sensitive the value of
$\Delta\mu/\mu$ determined from the model is to certain choices made. A
description of several performed tests is given below, while the outcome is summarized
in Table~\ref{table2}.

\begin{table*}
\begin{minipage}{126mm}
    \centering
    \caption{Resulting $\Delta\mu/\mu$ values and their statistical
uncertainties from various consistency tests
performed on the 12 velocity component absorption model as described in
Section~\ref{tests}. The $b \neq F[J]$ and
the $b= F[J]$ models are different in that the former has $b$ parameters tied in
all $J$ levels of corresponding velocity components whilst in the latter they
are only tied within every $J$ level of corresponding velocity components. The
last line of the table refers to a test, in which only the best exposures
(calibration-wise) were included in the spectrum. 
        }
    \begin{tabular}{c c c c c c}
    \hline\hline
   & $b$   & Transitions & n$_{\rm trans}$ & Components & $\Delta\mu/\mu$ [$\times 10^{-5}$]  \\[0.5ex]
    \hline
    & $\neq F[J]$   & all	 	& 32 	&all	& -0.68$\pm$2.70 \\ 
i   & $\neq F[J]$   & all		& 28 	&left	& 3.55$\pm$4.34\\
i   & $\neq F[J]$   & all	 	& 22 	&right	& -3.87$\pm$3.43\\ 
ii  & $\neq F[J]$   & $J = [0,1]$ 	& 11 	&all	&-0.86$\pm$3.50 \\ 
ii  & $\neq F[J]$   & $J = [2-5]$	& 21 	&all	&-0.46$\pm$4.06 \\ 
iii & $= F[J]$      & all	 	& 32 	&all	& -1.20$\pm$3.03 \\ 
iv  & $\neq F[J]$   & all	 	& 32 	&all	& 0.68$\pm$3.45\\   
\hline\hline

   \end{tabular}
        \label{table2}
\end{minipage}
\end{table*}

\begin{enumerate}
 \item{The extended structure of the H$_2$ absorption profile can be divided
into two parts. As it can be seen from velocity plots, the dividing line naturally falls at around
0~km~s$^{-1}$ (Fig.~\ref{fig1}). When only the left part of the profile, which
includes 4 H$_2$ velocity components, was used in the fitting, it resulted in
$\Delta\mu/\mu = (3.55 \pm 4.34_{\rm stat}) \times 10^{-5}$. The right part with
8 velocity components delivered $\Delta\mu/\mu = (-3.87\pm3.43_{\rm stat})\times 10^{-5}$. The larger
uncertainty of the left side result can be explained by the fact that several
of the fitted transitions were saturated, thus their centroids are determined
less accurately. Altogether, the two results are consistent within the combined
1.3-$\sigma$.}

\item{Instead of allowing all transitions to contribute to a single
$\Delta\mu/\mu$ value, within {\sevensize VPFIT} it is possible to fit a
different value of $\Delta\mu/\mu$ for each $J$-level or for $J$-levels grouped in some way. This
kind of test allows to quantify the relative contributions different $J$-levels
make to the final result. As not many transitions are available in the Q2348$-$011
absorber they were not investigated level by level. \citet{ubachs2007} have
suggested that H$_2$ transitions can be divided into a $J \in [0,1]$ set (cold
states) and a $J \ge 2$ set (warm states) to examine the impact of temperature:
due to the para-ortho distribution of H$_2$ the $J = 1$ state is significantly
populated even at low temperatures. A test was performed where only transitions from the cold states
or only those from the warm states were used to determine $\Delta\mu/\mu$. The
two values of $\Delta\mu/\mu$ match within the uncertainty. It means that the
two groups of transitions contribute similarly to the final combined value of
$\Delta\mu/\mu$.} 

\item{All previously described tests were performed on absorption models which
are based on the assumption that corresponding velocity components in all
transitions have the same $b$-parameter, independent of rotational quantum state
($b \neq F[J]$). This assumption is relaxed to test what its effect is on
$\Delta\mu/\mu$, i.e. in this run different $J$ transitions are allowed to have different $b$-parameters
in corresponding velocity components ($b = F[J]$). No significant discrepancy is
found between the values determined from the $b \neq F[J]$ and the $b = F[J]$
models.
}

 \item{The spectrum used in the analysis includes exposures from the older 2003
ESO-archive data set which do not have the individually taken Th-Ar calibration frames. They make up 4.50 hrs
compared to 19.25 hrs of more accurately calibrated data. In order to make sure that the
final result is not affected by the additional (possibly less accurate) data,
they were excluded in one of the fit optimizations. The newly determined value
is $\Delta\mu/\mu = $(0.68$\pm3.45_{\rm stat})\times 10^{-5}$. The result is
consistent with the one from the primary fit. The uncertainty is higher since the S/N of the spectrum used in
this test is lower.}
\end{enumerate}
The performed tests show that the final result on $\Delta\mu/\mu$ determined
from the 12 component model is not affected by a subclass of transitions,
velocity components or data.

\subsection{Systematic errors}

The statistical error of the fiducial result is larger than obtained in other
recent studies of H$_2$ studies (e.g., \citealt{malec2010}; \citealt{king2011}).
In those studies, systematic errors were found
to be smaller than, but comparable in magnitude to, the statistical errors.
Thus, while it is still important to consider systematic errors in the present
study, they are not expected to dominate the error budget.

Every echelle order included in the new Q2348$-$011 spectrum is calibrated using
about 10 Th-Ar lines. The effect of the errors of individual Th-Ar lines is
small (about 70 m~s$^{-1}$) and random, so it will average out when
many H$_2$ transitions spread over many orders are used.
\citet{murphy2007} have shown that possible systematic patterns in the
Th-Ar calibration residuals result in a deviation of 30 m~s$^{-1}$ at most. The
$K_i$ values used in the present analysis are in the range from $-0.015$ to
$0.018$. The effect of the systematic errors can be expressed via:
$\delta(\Delta\mu/\mu) = (\Delta v/c)/\Delta K_i$. A shift of 30 m~s$^{-1}$
translates to $\delta(\Delta\mu/\mu) = 0.3 \times 10 ^{-5}$ which is about
15 per cent of the statistical uncertainty. 

In a recent study targeted at the possible
miscalibration of the wavelength scale of the UVES spectrograph it is reported
that intra-order distortions of about 200 m~s$^{-1}$ in size may be present
\citep{withmore2010}. Their influence can be roughly estimated by
dividing the distortion by the square root of the number of the H$_2$
transitions included in the fit: 200$/\sqrt32 \approx 35$ m~s$^{-1}$.

\citet{malec2010} have considered additional sources of possible systematic
errors but they are all of the same order as the Th-Ar wavelength calibration
uncertainties and are not expected to have substantial impact on the result of
the present analysis. It is conservatively assumed that the systematic error is
at most 25 per cent of the statistical error of $\Delta\mu/\mu$ derived in the
present analysis.

\section{Result}

Based on the results provided in Table~\ref{table2},
\[\Delta\mu/\mu =
(-0.68\pm2.70_{\rm stat}\pm0.66_{\rm sys})
 \times 10^{-5} \]
is adopted as the fiducial result for the analysis of molecular hydrogen absorption in the Q2348$-$011
spectrum (or $\Delta\mu/\mu=(-0.68\pm2.78) \times 10^{-5}$, 
adding the uncertainties in quadrature). The result is consistent with
no change in $\mu$ at the $10^{-5}$ level.

The analysis of H$_2$ absorption in the Q2348$-$011 spectrum yields a result on
$\Delta\mu/\mu$ which is less tight compared to the constraints from previous
analyses (Ubachs et al. 2011). One of the major causes that led to the larger
uncertainty is the relatively low S/N ratio of the spectrum. The second cause is
that the number of lines is relatively small: 32 compared to e.g., 90 in the
sight-line of J2123$-$0050, which yields an order of magnitude more precise
result on $\Delta\mu/\mu$ (\citealt{malec2010}; \citealt{weerdenburg2011}).
Part of the H$_2$ spectrum in Q2348$-$011 is obscured
by the neutral hydrogen features of the additional strong DLA at $z \simeq
2.615$ making some of the relevant transitions
unavailable. The H$_2$ transitions falling at the short wavelength range of
the spectrum are especially useful in $\mu$-variation analysis, since they
exhibit the larger sensitivity coefficients. In the spectrum of Q2348$-$011 they
are not detected due to the H{\sevensize~I} absorption produced by the
additional DLA and another strong absorber at $z\simeq 2.93$. However, the
present analysis demonstrates that a complex absorption structure can be
successfully modelled by means of the comprehensive fitting method.

\section{Conclusion}

Molecular hydrogen features present at redshift $z_{\rm abs} \simeq 2.426$ in
the line-of-sight to the quasar Q2348$-$011 were analysed to detect
a possible variation of the proton-to-electron mass ratio on a cosmological
time-scale. The constraint derived in the analysis is
$\Delta\mu/\mu =(-0.68\pm2.78) \times 10^{-5}$, showing no
indication that $\mu$ in this particular absorber is different from $\mu$
measured in the laboratory. Although being less accurate than other recent H$_2$
constraints on $\Delta\mu/\mu$, the result is consistent
with those results, which show that
$\left|\Delta\mu/\mu\right| < 1\times10^{-5}$ at $z=2$--3.

In the sample of known high-redshift H$_2$ absorbers, the structure of the
absorber in the Q2348$-$011 spectrum is the most complex as it has 7 visually distinguishable
velocity components, whilst 12 are justified statistically. The present analysis
shows the applicability of the comprehensive fitting method in a case of such a complex structure. However, it
also shows that in order to achieve a competitive result on $\Delta\mu/\mu$, the
H$_2$ absorber/spectrum selected for analysis must obey several conditions among
which are no occurrence of strong additional absorbers and a sufficiently high
signal-to-noise ratio.

\section*{Acknowledgments}

This work is based on observations carried out at the European
Southern Observatory (ESO) under programme ID 79.A-0404(A) (PI Ubachs), with the UVES
spectrograph installed at the Kueyen UT2 on Cerro Paranal, Chile. Additional
data from the ESO-archive were used from programme ID 072.A-0346(A). JB
would like to acknowledge F.~van~Weerdenburg and A.~Malec, for assistance with
using {\sevensize VPFIT} and fruitful discussions. MTM thanks
the Australian Research Council for a QEII Research Fellowship (DP0877998). WU
acknowledges support from the Netherlands Foundation for Fundamental Research of
Matter (FOM).

\label{lastpage}


\begin{thebibliography}{99}

\bibitem[\protect\citeauthoryear{Bailly et al.}{2010}]{bailly} Bailly D.,
Salumbides E.~J., Vervloet M., Ubachs W., 2010,
 Mol. Phys., 108, 827

\bibitem[\protect\citeauthoryear{Ivanov et al.}{2008}]{ivanov2008} Ivanov T.~I.,
Roudjane M., Vieitez M.~O., de Lange C.~A., Tchang-Brillet W.-{\"U}.~L., Ubachs
W.,
 2008, Phys. Rev. Lett., 100, 093007
\bibitem[\protect\citeauthoryear{Ivanov et al.}{2010}]{ivanov2010} Ivanov T.~I.,
Dickenson G.~D., Roudjane M., Oliveira N.~De, Joyeux D., Nahon L.,
Tchang-Brillet W.-{\"U}.~L., Ubachs, W., 2010
Mol. Phys., 108, 771
\bibitem[\protect\citeauthoryear{King et al.}{2008}]{king2008} King J.~A., Webb
J.~K., 
Murphy M.~T., Carswell R.~F., 2008, Phys. Rev. Lett., 101, 251304

\bibitem[\protect\citeauthoryear{King et al.}{2011}]{king2011} King J.~A.,
Murphy 
M.~T., Ubachs W., Webb J.~K., 2011, MNRAS, 417, 3010

\bibitem[\protect\citeauthoryear{Ledoux et al.}{2003}]{ledoux2003} Ledoux C.,
Petitjean P., Srianand R., 2003, MNRAS, 346, 209

\bibitem[\protect\citeauthoryear{Ledoux et al.}{2006}]{ledoux2006} Ledoux C.,
Petitjean P., Fynbo J.~P.~U., Moller P., Srianand R., 2006, A$\&$A, 457, 71

\bibitem[\protect\citeauthoryear{Malec et al.}{2010}]{malec2010} 
Malec A.~L., et al., 2010, MNRAS, 403, 1541 

\bibitem[\protect\citeauthoryear{Meshkov et al.}{2006}]{meshkov2006} Meshkov
V.~V., Stolyarov A.~V., Ivanchik A., Varshalovich D.~A., 2006, JETP Lett., 83,
303

\bibitem[\protect\citeauthoryear{Murphy et al.}{2007}]{murphy2007} Murphy
M.~T., Tzanavaris P., Webb J.~K., Lovis C., 2007, MNRAS, 378, 221

\bibitem[\protect\citeauthoryear{Murphy, Webb~$\&$~Flambaum}{2008}]{murphy2008}
Murphy M.~T.,
Webb J.~K., Flambaum V.~V., 2008, MNRAS, 384, 1053

\bibitem[\protect\citeauthoryear{Noterdaeme et al.}{2007}]{noterdaeme2007}
Noterdaeme P., Petitjean P., Srianand R., Ledoux C., 2007, A$\&$A, 469, 425

\bibitem[\protect\citeauthoryear{Petitjean et al.}{2006}]{petitjean2006}
Petitjean P., Ledoux C., Noterdaeme  P., Srianand R., 2006, A$\&$A, 456, L9

\bibitem[\protect\citeauthoryear{Philip et al.}{2004}]{philip2004} Philip J.,
Sprengers J.~P., Cacciani P., de Lange C.~A., Ubachs W., 2004, Applied Physics
B: Lasers and Optics, 78, 737

\bibitem[\protect\citeauthoryear{Reinhold et al.}{2006}]{reinhold2006} 	Reinhold
E., Buning R., Hollenstein U., Ivanchik A., Petitjean P., Ubachs W.,
 2006, Phys. Rev. Lett., 96, 151101

\bibitem[\protect\citeauthoryear{Salumbides et al.}{2008}]{salumbides} 
Salumbides E.~J., Bailly D., Khramov A., 
Wolf A.~L., Eikema K.~S.~E., Vervloet M., Ubachs W., 2008, Phys. Rev. Lett.,
101, 
223001

\bibitem[\protect\citeauthoryear{Srianand et al.}{2005}]{srianand2005} Srianand
R., 
Petitjean P., Ledoux C., Ferland G., Shaw G., 2005, MNRAS, 362, 549


\bibitem[\protect\citeauthoryear{Thompson}{1975}]{thompson}  Thompson R.,  1975,
Astroph. Lett., 16, 3

\bibitem[\protect\citeauthoryear{Thompson et 
al.}{2009}]{thompson2009} Thompson R.~I., et al., 2009, ApJ, 703, 
1648 

\bibitem[\protect\citeauthoryear{Ubachs et al.}{2007}]{ubachs2007} Ubachs W.,
Buning R., Eikema K.~S.~E., Reinhold E., 2007, J. Mol. Spectr., 241, 155

\bibitem[\protect\citeauthoryear{Ubachs et al.}{2011}]{ubachs2011} Ubachs W.,
Bagdonaite J., Murphy M.~T., Buning R., Kaper L., 2011, in Martins C., Molaro
P., eds., Astrophys. Space
Science, From Varying Couplings to Fundamental Physics, p. 125 

\bibitem[\protect\citeauthoryear{van Weerdenburg et al.}{2011}]{weerdenburg2011}
van 
Weerdenburg F., Murphy M.~T., Malec A.~L., Kaper L., Ubachs W., 2011, Phys. Rev.
Lett., 106, 180802  

\bibitem[\protect\citeauthoryear{Varshalovich \&
Levshakov}{1993}]{varshlovich1993} Varshalovich D.~A., Levshakov S.~A.,
1993, Soviet Journal of Experimental and Theoretical Physics Letters, 58, 237 

\bibitem[\protect\citeauthoryear{Wendt~$\&$~Molaro}{2010}]{wendt2010} Wendt M.,
Molaro P., 2010,  A$\&$A, 526, A96


\bibitem[\protect\citeauthoryear{Wendt~$\&$~Reimers}{2008}]{wendt2008} Wendt M.,
Reimers D., 2008,  Eur. J. Phys. D Spec. Top., 163, 197

\bibitem[\protect\citeauthoryear{Whitmore,
Murphy~$\&$~Griest}{2010}]{withmore2010} Whitmore
J.~B., 
Murphy M.~T., Griest K., 2010, ApJ., 723, 89 



\end{thebibliography}
\end{document}